# Accelerated Design of Chalcogenide Glasses through Interpretable Machine Learning for Composition–Property Relationships

Sayam Singla[1], Sajid Mannan[2], Mohd Zaki[2], N.M. Anoop Krishnan[2,3]*

[1]Department of Textile and Fibre Engineering, Indian Institute of Technology Delhi, Hauz Khas, New Delhi, 110016, India

[2]Department of Civil Engineering, Indian Institute of Technology Delhi, Hauz Khas, New Delhi, 110016, India

[3]Yardi School of Artificial Intelligence, Indian Institute of Technology Delhi, Hauz Khas, New Delhi, 110016, India

*Corresponding author: krishnan@iitd.ac.in

**Abstract**

Chalcogenide glasses possess several outstanding properties that enable several ground-breaking applications, such as optical discs, infrared cameras, and thermal imaging systems. Despite the ubiquitous usage of these glasses, the composition–property relationships in these materials remain poorly understood. Here, we use a large experimental dataset comprising ~24000 glass compositions made of 51 distinct elements from the periodic table to develop machine learning models for predicting 12 properties, namely, annealing point, bulk modulus, density, Vickers hardness, Littleton point, Young's modulus, shear modulus, softening point, thermal expansion coefficient, glass transition temperature, liquidus temperature, and refractive index. These models, by far, are the largest for chalcogenide glasses. Further, we use SHAP, a game theory-based algorithm, to interpret the output of machine learning algorithms by analyzing the contributions of each element towards the model's prediction of a property. This provides a powerful tool for experimentalists to interpret the model's prediction and hence design new glass compositions with targeted properties. Finally, using the models, we develop several glass-selection charts that can potentially aid in the rational design of novel chalcogenide glasses for various applications.

## Introduction

Chalcogenide glasses (ChGs) are a family of glasses that contains one or more chalcogen elements, namely, Sulphur (S), Selenium (Se), or Tellurium (Te), in combination with other elements from Group IV, V, and VI of the periodic table except for oxygen [1], [2]. ChGs have been considered as a typical family of non-oxide glasses because of the huge difference in covalent bonding nature and its physical, mechanical, and optical properties as compared to oxide glasses [3]. ChGs possess several outstanding properties, such as small band gap energy ($E_g$ ~ 1-3 eV), giving rise to their semiconducting properties, photosensitivity, and fast ionic conductivity (>$10^{-3}$ Scm$^{-1}$), which leads to a wide range of applications [4]. Due to these reasons, they are widely used in optical discs, infrared cameras, and thermal imaging systems for automobile navigation, as a waveguide in optical circuits and as a solid electrolyte in all solid-state batteries [5], [6], some of which are active areas of research and development.

Although the glass forming ability of chalcogens has been known for several decades [7], compared to oxide glasses, they are still emerging and under-explored. One of the major reasons for the limited understanding is that the composition–property relationships in these glasses are still elusive. Traditionally, the Edisonian trial-error method was used to develop new glasses with particular property values based on experience and domain knowledge, but this approach is resource intensive and time consuming. Recently, researchers have been actively using machine learning algorithms to develop a composition property model with the available curated databases for predicting properties such as Young's modulus, glass transition temperature, and refractive index of oxide glasses [8]–[19]. These approaches combined with artificial intelligence-based information extraction from the literature have proven to significantly accelerate materials design and discovery by enabling automated development of databases and knowledgebases from the literature [17], [18], [20], [21]. In addition, data-driven modelling has proved to be quite helpful in understanding the composition–property relationships and, thereby, enhancing the design of new glasses [22], [23].

However, there have been limited studies on predicting and interpreting the properties of chalcogenide glasses. For instance, recently, Mastelini et al. [23] trained different ML models such as random forest, K-nearest neighbours, neural network and classification and regression trees (CART) for predicting three different properties of chalcogenide glasses, namely, glass transition temperature, refractive index, and thermal expansion coefficient using the dataset collected from Sciglass. They used 456 and 7620 unique glass compositions with 1 to 6 different elements in each glass for predicting refractive index and glass transition temperature, respectively. To the best of the authors' knowledge, this is the only prior work where ML has been used for modeling chalcogenide glasses. However, several other important physical and mechanical properties of chalcogenide glasses remain to be modeled. Moreover, the use of these properties to design new glass compositions in an accelerated fashion needs to be explored and elucidated.

To address this, here we develop ML models for twelve properties of chalcogenide glasses such as annealing point (*AP*), bulk modulus (*K*), density (*ρ*), Vickers hardness (*$H_v$*), Littleton point (*LP*), Young's modulus (*E*), shear modulus (*G*), softening point (*SP*), thermal expansion coefficient (*TEC*), glass transition temperature ($T_g$), liquidus temperature ($T_L$), and refractive index ($n_d$) by considering up to 51 distinct elements from the periodic table. These models are the largest ones for chalcogenide glass reported in the literature. To address the black box nature of these ML models, we use the SHAP (SHapley Additive exPlanations) algorithm based on game theory to understand the interdependencies of elements towards different property values. It is well-known that the components in a glass composition cannot be treated independently [2]. Thus, we have plotted the interaction value plot and selection chart using the developed ML models to see the correlation of different features and their dependency on the property value and design new glasses with enhanced property value, respectively. Altogether, this study will help in accelerating the design and discovery of new chalcogenide glasses with tailored properties.

**Methodology**
*Data Collection and Data Processing*
Here we collected the data from SciGlass and INTERGLAD V7 databases, well-known databases for composition-property value. These two datasets together contain about

350,000 glass compositions from different sources like research papers, patents, handbooks, etc. We extracted compositions and properties of chalcogenide glasses from these databases to train ML models for predicting their property value. We follow the cleaning methodology mentioned by Mastelini et al. (2021)[24]. First, we removed all the glasses which contained any amount of oxygen, nitrogen, fluorine, chlorine, bromine, iodine, gold, silver, platinum, and palladium and considered only those glasses which contained at least one of the non-zero amount of sulfur, selenium, or tellurium. The extracted data were converted into atomic mol% using stoichiometric analysis. For example, for a glass composition $20(As_2S_3).80(As_2Te_3)$, the input feature would be 40, 12 and 48 corresponding to constituting elements As, S, and Te, respectively, and the output would be the property value of interest. After conversion, we removed all the glasses whose composition percentage doesn't add up to 100%, and duplicate entries were merged by taking the mean of each property corresponding to duplicate entries while dropping the outliers which have a value beyond the interval of ± 3 times the standard deviation of the corresponding property values [8].

Note that only those glass compositions were selected for which the property value was in the interval of ± 3 times the standard deviation of all the property values. Also, only those elements were taken that are present in at least 20 glass compositions. In this work, we report machine learning models to predict the 12 important properties of chalcogenide glasses, which are as follows, annealing point (*AP*), bulk modulus (*K*), density (*ρ*), Vickers hardness ($H_v$), Littleton point (*LP*), Young's modulus (*E*), shear modulus (*G*), softening point (*SP*), thermal expansion coefficient (*TEC*), glass transition temperature ($T_g$), liquidus temperature ($T_L$), and refractive index ($n_d$). Note that the names and abbreviations are reiterated for the sake of complete description. The cleaned dataset was further divided into training and test sets, having 80 percent and 20 percent of all the data points for a given property. The training set was subjected to 4-fold cross-validation, and the best model was selected by training it on 3-folds and evaluating it on the 4$^{th}$ fold of the training set. This approach was repeated with all folds as the evaluation fold and enabled hyperparametric optimization. Finally, the model performance was evaluated on the unseen 20% test dataset that was kept separate. This method of processing was consistent for all 12 properties. All the codes used for training the models are provided in https://github.com/M3RG-IITD/chgs-ai.

*Model Development*
Despite the fact that we train different ML models for each of the 12 qualities of interest, the XGBoost model [25] surpasses all other ML models for every property. This observation is consistent with our previous study as well, where we trained 25 property models for oxide glasses [25]. The composition of glass in mol% is used as the input features for all 12 properties, and the output of the model is the accompanying property values in their respective units. Using the squared error metric as the loss function for the optimizer, two types of boosters have been used: gbtree and Dart.

In addition, hyperparameter tuning is performed with the aid of the optuna package [26], which offers a quick sampling and pruning approach to optimising the objective function by providing the validation score given a given set of hyperparameters. Independent (TPE [27]) and relational (CMA-ES [28], GP-BO [29]) sampling techniques are used to study new trials. It uses pruning techniques like Asynchronous Successive Halving [30] to eliminate futile trials (ASHA). It also has an interface for incorporating one's own sampling and pruning methods. The final product is a trained machine-learning model with optimal hyperparameter settings.

**Results and discussion**
*Data Visualizations*
To visualize all the compositions used in this work, we first use an unsupervised machine learning algorithm of k-means to cluster the compositions into 30 clusters. The similar compositions were assigned to the same group. These compositions are then projected in two dimensions using t-SNE embeddings. Figure 1(a) shows, a two-dimensional t-SNE plot of all the glass compositions color-coded into clusters according to the outcomes of k-means clustering. The names of the three most prevalent glass components appear next to the 13 most prominent clusters, while the remaining clusters are grouped together as "miscellaneous."

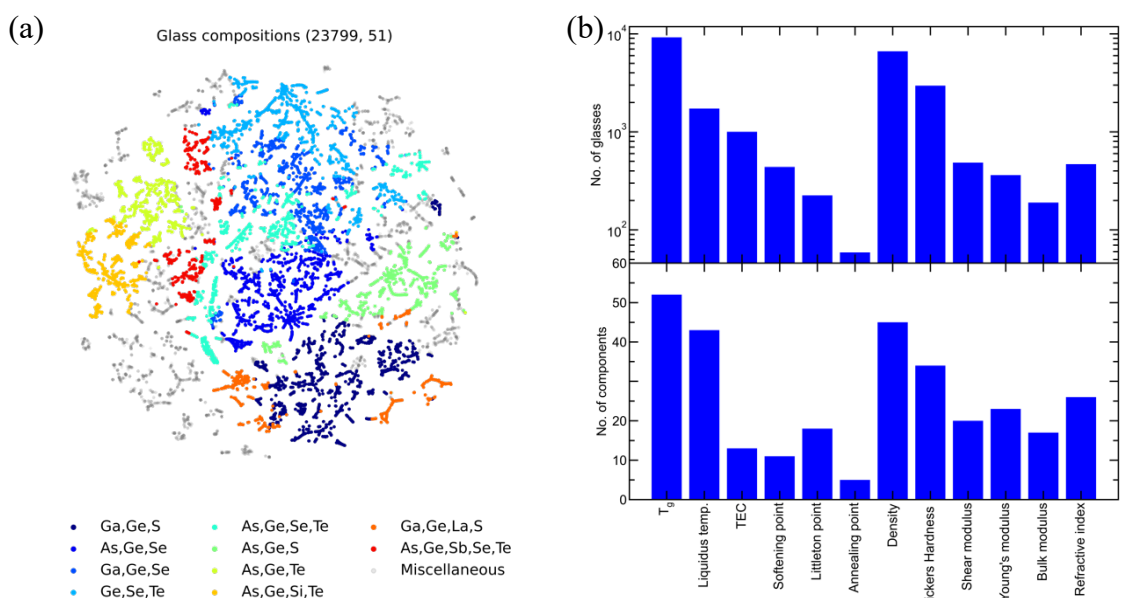

**Figure 1.** (a) represents all the glass compositions used in this study embedded in a two-dimensional t-SNE plot. The influential families were marked according to the color in legends. (b) Bar chart displaying the total number of glass compositions and components for given properties.

Figure 1(b) consists of two subplots in which the upper subplot shows the bar plot of the number of glass compositions for each of the properties. For most of the properties, the number of glass samples is less than 1000, and the glass transition temperature contains the highest number of data points with around 10,000. Instead of removing the Annealing point that includes the least number of samples with 60 points, we used it to show the extent of interpretability of the models developed in this work. The lower subplot indicates the number of components for each property. Properties like glass transition temperature with a maximum of 51 elements and density show higher values as they are standard properties, and most of the research papers mention glass properties. Annealing point contains the least number with 4 distinct elements. Overall, the number of features varies widely, ranging from 4 elements for the Annealing point to 51 elements for the glass transition temperature. Detailed visualization of the datasets corresponding to all the properties, including the distribution of the property values, histogram of the number of glasses with *n*-components, and the number of glasses with any given element as a component for a property are given in the Supplementary material.

*Model predictions*

After training the different models using the steps described in the methodology section, we evaluated it on the test dataset that was kept hidden during the training and validation phase. The hyperparameters associated with the training are included in the Supplementary material. Fig. 2 and 3 show the comparisons between the measured and predicted properties of ChGs. The points belonging to training and test sets are shown with blue and pink color respectively. The performance of trained models is shown as $R^2$ scores within the respective subfigures. Further, the straight line at 45 degrees is shown to give a qualitative idea of how many points are closer to this line implying the goodness of the fit.

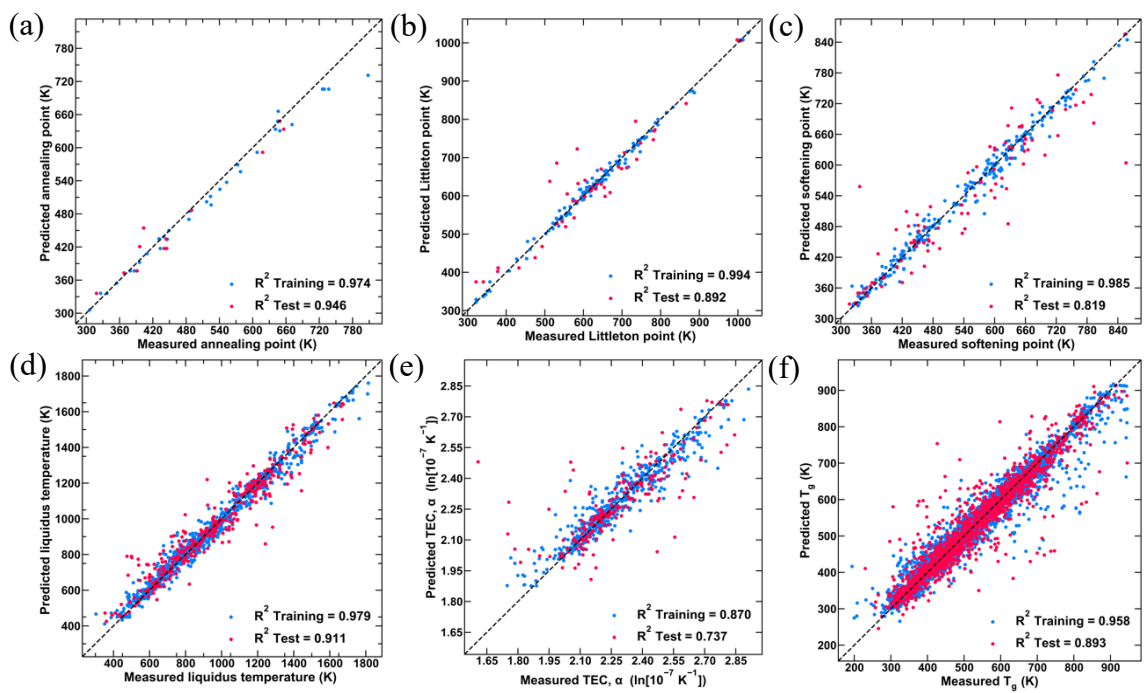

**Figure 2.** Predicted vs experimental values for physical properties, namely, (a) annealing point (*AP*), (b) Littleton point (*LP*), (c) softening point (*SP*), (d) liquidus temperature (*$T_L$*), (e) thermal expansion coefficient (*TEC*), and (f) glass transition temperature ($T_g$).

Figure 2(a-f) shows predicted vs measured values for 6 physical properties, namely, *AP, LP, SP, $T_L$, TEC,* and *$T_g$,* based on the optimized XGBoost models. For most properties, we observe that the $R^2$ values are high, suggesting that (a)the models can predict the properties reasonably, and (b) compositional variation mainly governs these properties, and no additional information or constraint is required to modulate the predicted variable. However, for some properties, such as the thermal expansion coefficient and the softening point have a slightly lower value of $R^2$ score, suggesting that these properties might be influenced by other parameters like measurement techniques. This could also be attributed to the noise in the measurement of these properties, which might be higher than the other properties mentioned.

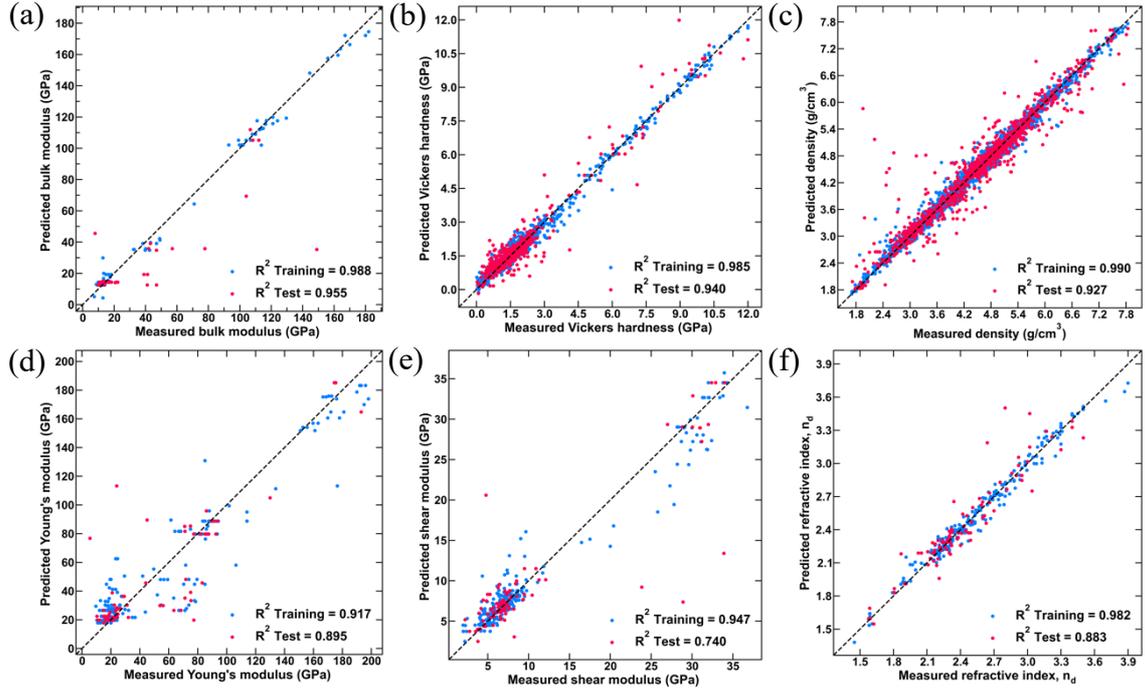

**Figure 3.** Predicted vs experimental values for mechanical properties, i.e., (a) bulk modulus ($K$), (b) vickers's hardness ($H_v$), (c) density ($\rho$), (d) young's modulus ($E$), (e) shear modulus ($G$), (f) refractive index ($n_d$).

Figure 3(a-f) shows predicted vs measured values for five mechanical and one optical property obtained using the trained XGBoost model. It has been observed that $R^2$ scores are (> 0.9) for the $K$, $E$, and $\rho$, which implies these properties can be predominantly compositional dependent. However, the $n_d$ and $G$ have slightly lesser $R^2$ scores, and this can possibly be because of the less data points available for the model training. Further, especially in the case of shear modulus, which is generally measured indirectly, we observe that the data is highly clustered around two values, one lower and one higher. Such non-uniform distribution of the data could also be the cause for the poor performance of the ML model.

*Elucidating the composition–property relationship*
Now, we apply the Shapley additive explanations (SHAP), a game-theoretic technique, to explain the compositional control of the properties. The contribution of each glass component to the final forecast for each composition is measured by SHAP in terms of the characteristic's mean value for each composition. This enables both qualitative and quantitative interpretations of the role played by the glass components in controlling a certain quality. For each of the 12 properties—$AP$, $K$, $\rho$, $H_v$, $LP$, $E$, $G$, $SP$, $TEC$, $T_g$, $T_L$, $n_d$—the SHAP values for the top 5 glass components are displayed in Figs. 4 and 5. The components are arranged from bottom to top with an upward trend in the mean absolute SHAP values. Note that the SHAP plots essentially reveal the influence of a component towards increasing or decreasing a property value and the SHAP value, given in the $x$-axis, provides the quantitative increment that a given component will provide to the predicted property value. The results of SHAP value can be used to determine the following: Firstly, components that affect property value positively or negatively; secondly, components that have conflicting effects depending on the other components in the glasses; and lastly, the precise percentage increase in predicted property value for a given compositional value of the component. We have divided all 12 properties into physical, mechanical, and optical groups based on their application for better understanding.

*Physical properties*

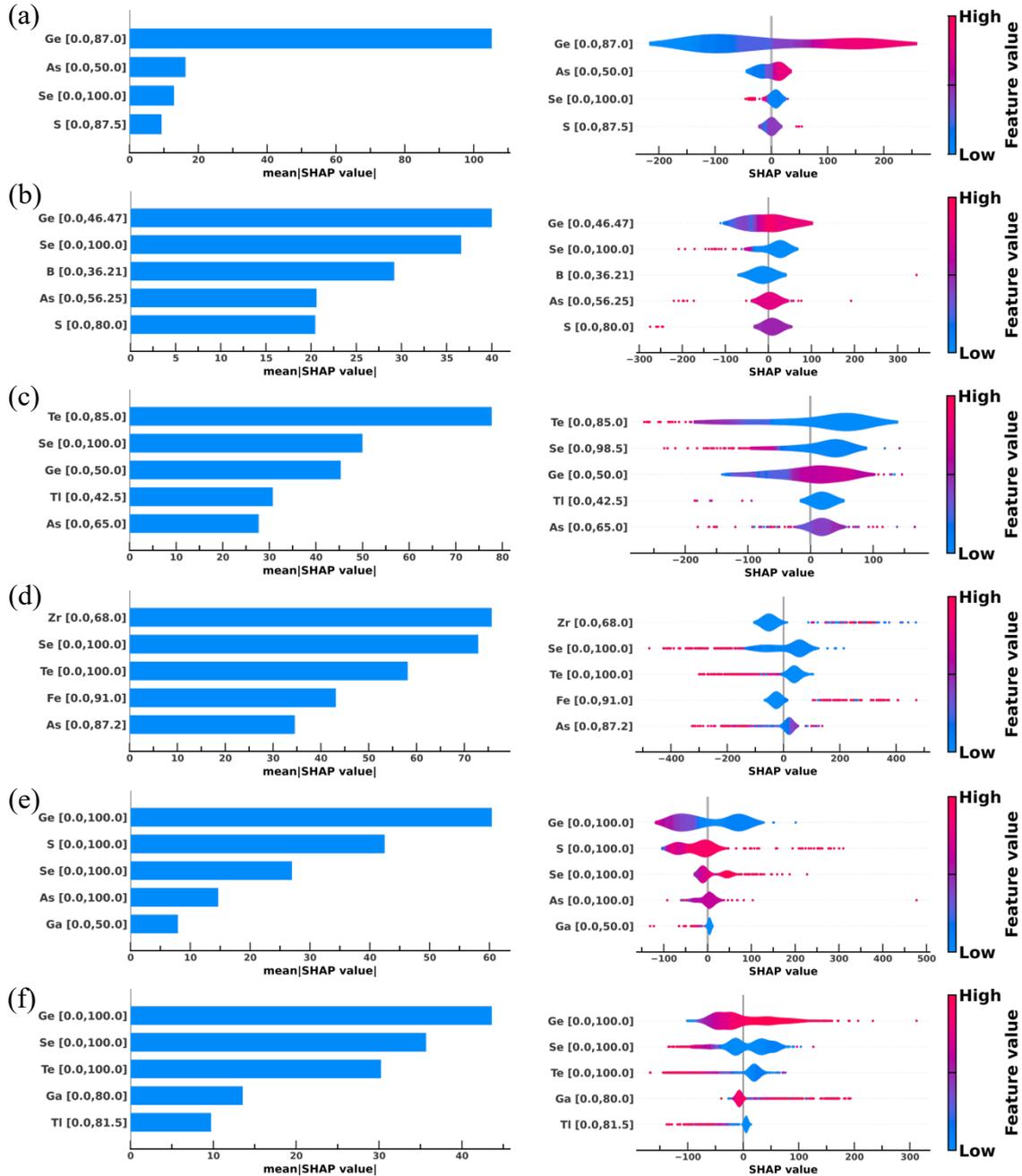

**Figure 4.** Interpretation of composition–property relation for physical properties using SHAP bar and violin plots. (a) annealing point (*AP*), (b) Littleton point (*LP*), (c) softening point (*SP*), (d) liquidus temperature ($T_L$), (e) thermal expansion coefficient (*TEC*), and (f) glass transition temperature ($T_g$).

Figure 4(a) shows that Germanium (Ge) enhance the *AP*, whereas Selenium (Se) impacts negatively and thus reduces the property value. Figure 4(b) shows elements like Ge, Se, Boron (B), and Arsenic (As) show higher percentages of value in the determination of *LP* and impact positively. However, Se has a negative impact, whereas B has mixed effects. Figure 4(c) shows Tellurium (Te), Se and Ge as the major contributors amongst the elements and has a positive effect on *SP*, but Te and Se have a negative impact on *SP* value, whereas Ge has mixed effects. Figure 4(d) shows elements like Zirconium (Zr), Se, Te, and Iron (Fe) show higher percentages of value in the determination of $T_L$. Se and Te show a negative impact, whereas other elements mentioned have mixed effects. Note that the addition of elements such as Se, Te, and S generally reduces the degree of connectivity in the glass matrix by modifying the three dimensional network that is usually formed by Ge and Gallium (Ga) [31], [32] in the glass structure. The rings formed by the glasses with the addition of these elements are easier to break, and hence the addition of elements such as Se, Te, and S results in lower values of physical properties such as *AP*, *LP*, *SP,* and $T_L$. This is consistent with the results obtained in the SHAP plots.

Figure 4 (e) shows that Se and Sulfur (S) negatively impact *TEC* value, whereas Ge has a positive impact. Figure 4(f) depicts the elements like Ge, Se, Te, and Ga are, the elements showing higher percentages of value in the determination of $T_g$ value. Similar to other physical properties, Se negatively impacts $T_g$, while Ge positively impacts it. However, suppose we increase their content beyond saturation. In that case, it will form a homopolar Ge-Ge bond instead of other elements, and thus it will reduce the Tg [33], while other elements mentioned have mixed effects. The experimentalists have also noticed that as more Ge is added to the glass system, it develops a two-dimensional character held by van der Waals forces, leading to an increase in $T_g$ [34]. It is interesting to observe from Fig. (e) that the effect of Ge and Se on *TEC* is opposite compared to their effect on other properties. This is consistent with the observation in oxide glasses that highly polymerized glasses will result in lower TEC while that with lower polymerization can have larger TEC values [35], [36].

*Mechanical and optical properties*

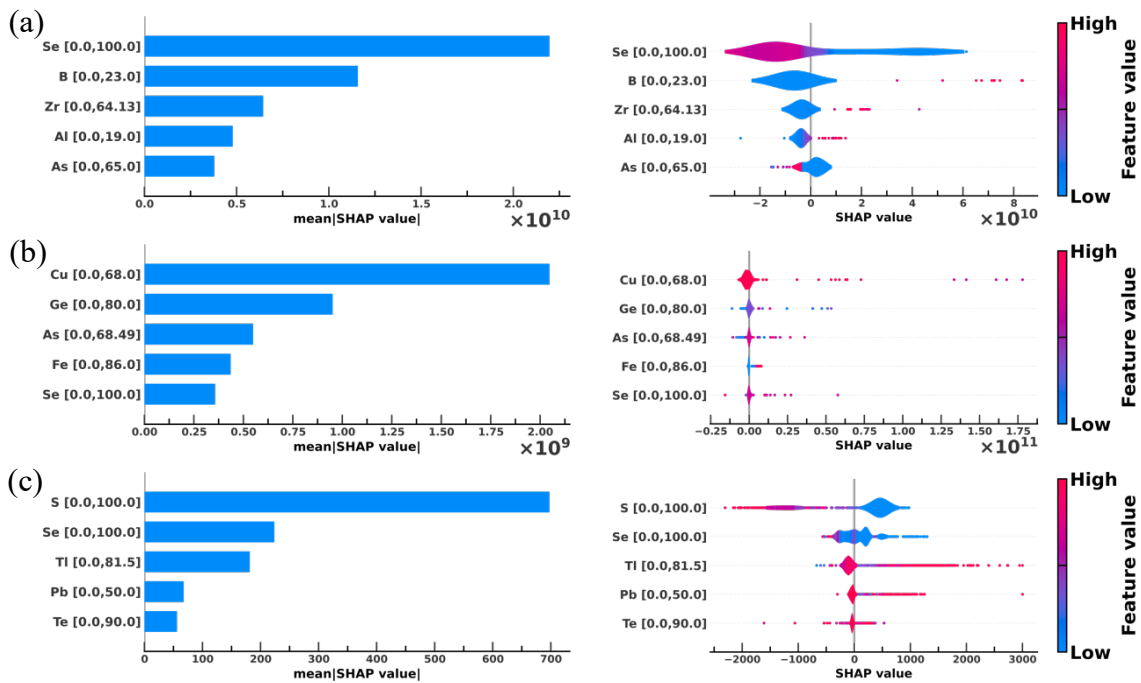

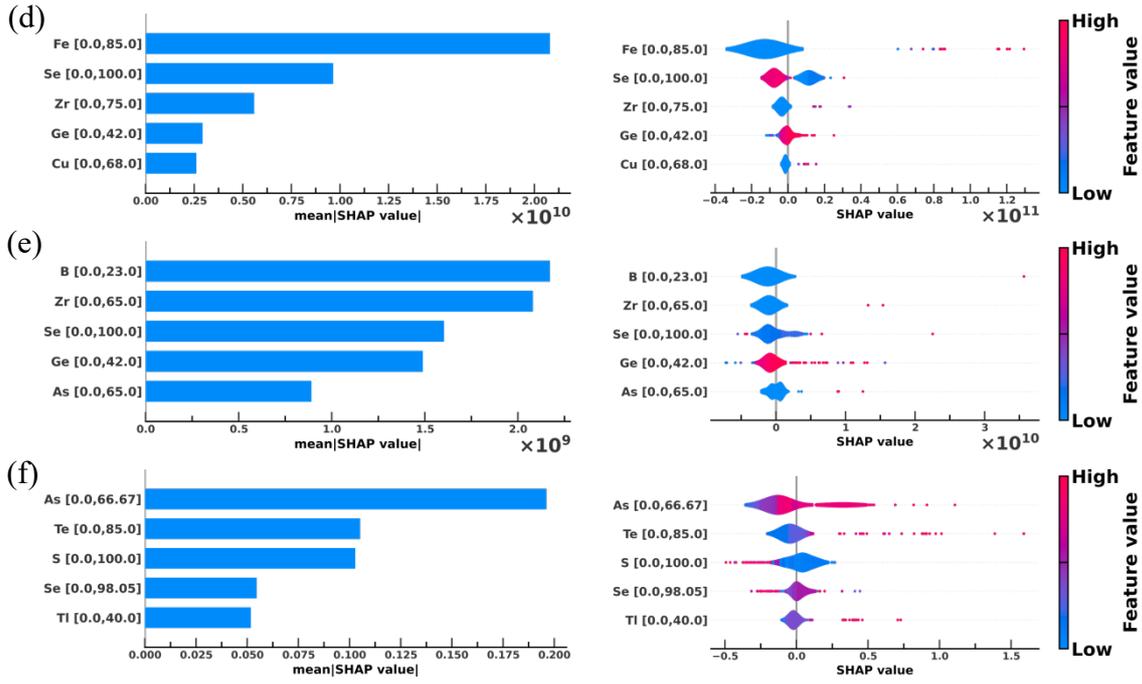

**Figure 5.** Interpretation of composition–property relation for mechanical properties using SHAP bar and violin plots. (a) bulk modulus ($K$), (b) Vickers hardness ($H_v$), (c) density ($\rho$), (d) young's modulus ($E$), (e) shear modulus ($G$), (f) refractive index ($n_d$).

Figure 5(a) shows the top five elements that govern the bulk modulus of ChGs. Se and As have a negative impact on $K$, Al has a positive impact and B and Ge have mixed effects. Figure 5(b) shows that high percentages of Copper (Cu), Ge, and As increase the model prediction of $H_v$. Higher Cu concentration positively impacts $H_v$, as suggested by previous research where Cu shows linear dependence with $H_v$ [37]. Ge also shows a positive impact, whereas As shows mixed effects on $H_v$. Figure 5(c) shows the prominent elements governing the density of CHGs. It is observed that Sb, Pb, and Tl impact the model prediction of $\rho$ positively. This could be possibly attributed to the relatively heavier atomic mass of the network formers, which dominates the density change while forming bonds with other elements like Silicon (Si), Phosphorous (P) and As [38]. It also shows that the higher S and Se content negatively impact the $\rho$, whereas Tl and Ge have a mixed effect on $\rho$. It is experimentally proven that higher S content negatively impacts and shows linear dependence with $\rho$ [39]. Figure 5(d) shows the compositional control of $E$ value. Higher values of both Ge and Fe lead to higher $E$ values, and this could be possible because of the increase in tetrahedral structure as $GeSe_2$ in a glass matrix, which eventually leads to the formation of a rigid network structure of glass, and thus it enhances the elastic properties and shear modulus[40], Se has mixed effects whereas Zr content can reduce the $\rho$.

Similarly, with $H_v$, most of the chalcogenide glasses generally have a low $E$ value. Figure 5(e) B and Zr negatively impact $G$, and Se, Ge, and As show mixed effects. Figure 5(f) shows that Fe and Ge positively impact the $n_d$, whereas S and Se reduce the model output, and B shows a mixed effect. $n_d$ is mainly governed by the electron shell polarizability and packing density of the network structure. Ge being a network modifier, easily forms a chain with other elements and enhances the $n_d$ [41]. These SHAP plots tells only about the individual contribution of element towards the property predictions, however, the overall property values is also governed by the interaction among constituent elements which can be understood using the SHAP interaction plots as shown in Figure 6 and 7.

*Interaction value plot*

Understanding the composition-property relationships and the interactions among components in governing the properties of ChGs is a crucial aspect of glass science. The examples of such interactions include the mixed-modifier effects and the boron anomaly [3] [14]. Knowing how two elements interact with each other in a multicomponent glass only tells the partial governance of properties. Understanding these detailed correlations requires high-precision experimentation, such as magic angle spinning nuclear magnetic resonance studies, or advanced computational simulations, such as density functional theory (DFT) modelling of the glass structure [8]. Here, using the SHAP interaction values (Fig. 6 and 7), we seek to ascertain the link between each input component and the attributes. The interaction plots show how different elements interact with each other to govern the properties of ChGs.

*Physical properties*

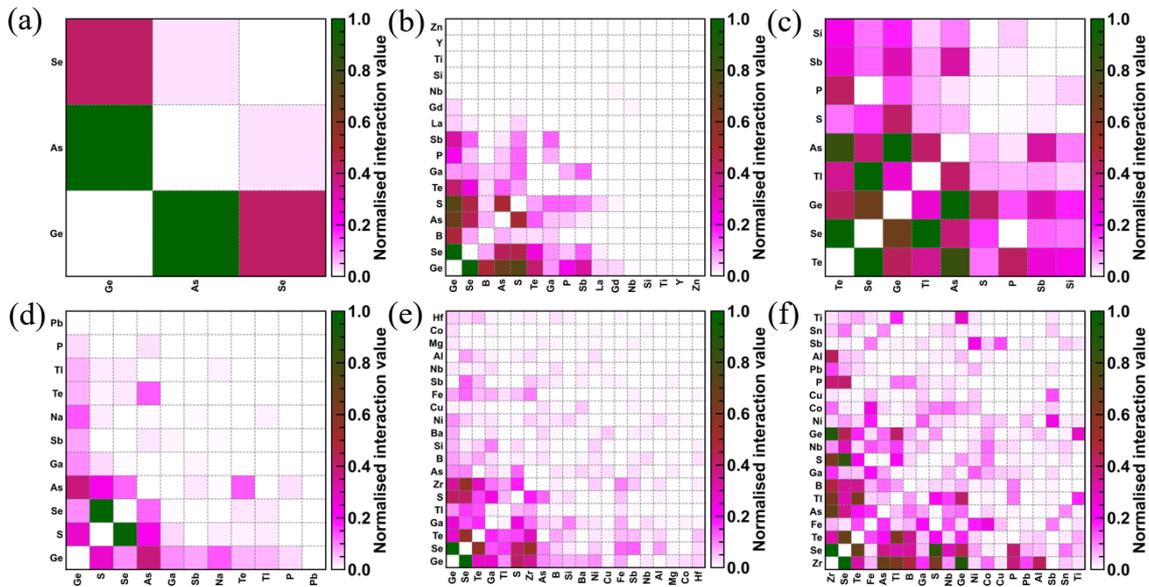

**Figure 6**. Interpretation of interdependency among input features for physical properties using Heat map (a) annealing point (*AP*), (b) Littleton point (*LP*), (c) softening point (*SP*), (d) liquidus temperature (*$T_L$*), (e) thermal expansion coefficient (*TEC*), and (f) glass transition temperature ($T_g$).

Figure 6(a) shows the pair of network formers Ge-As governing the AP of glasses. It indicates Ge has the highest interaction value with other elements while controlling the annealing point. Se, As, and S have higher interaction values for *LP* with other elements. Figure 6(c) indicates Ge has the highest interaction value with other elements. Pairs like Ge-As and Ge-Se show higher interactions among all pairs for *SP*. Figure 6(d) indicates elements like Zr, Se, Te, and As having higher interaction values with other elements. Major pairs with high interaction values for *$T_L$* are Zr-Se, Zr-Ge and S-Se. Figure 6(e) indicates Ge, S, Se and, As, with S-Se, Ge-As and Ge-S being the three combinations of elements with the highest interaction value for *TEC*. Figure 6(f) indicates elements like Ge, Se, Te, and S having higher interaction values with other elements. Pairs like Ge-Se, Se-Te, and Se-Zr show higher interaction among all pairs for $n_d$. In addition, the more populated the interaction value plot, the more complex the interactions in the system.

*Mechanical and Optical properties*

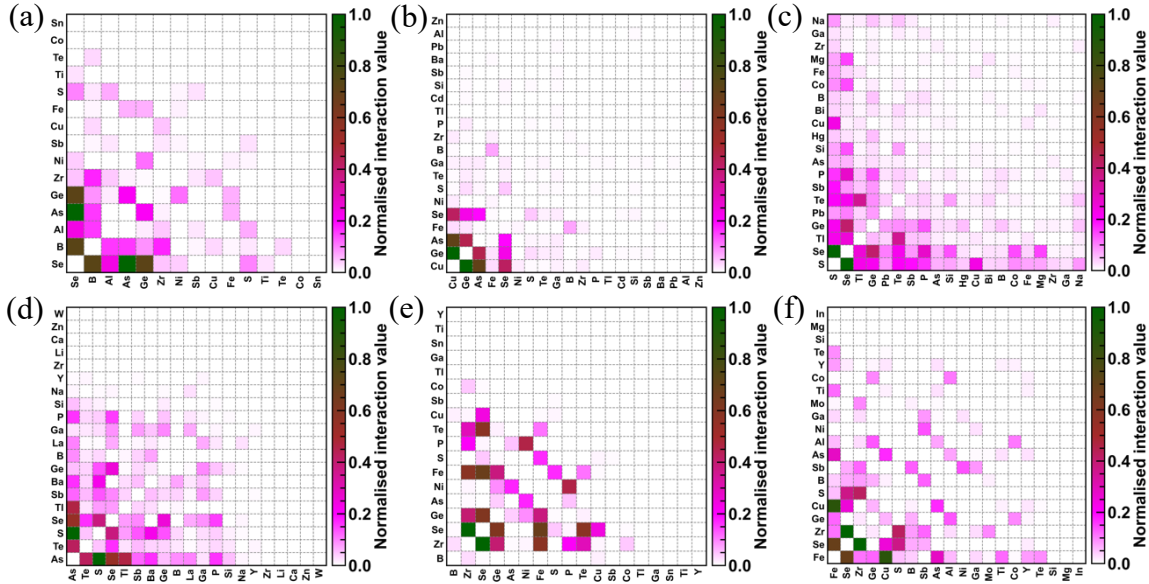

**Figure 7.** Interpretation of interdependency among input features for Mechanical properties using a Heat map (a) bulk modulus (*K*), (b) Vickers hardness ($H_v$), (c) density ($\rho$), (d) young's modulus (*E*), (e) shear modulus (*G*), (f) refractive index ($n_d$).

Figure 7(a) indicates Se, B, As and Ge, with Se-As, Se-B and Se-Ge are the three combinations of elements with the highest interaction value showing Se being the common element in pairs showing its major importance in controlling *K*. Figure 7(b) shows elements like Cu, Ge, As and Se having higher interaction-value with other elements. As shown in Fig 4(d), most of the chalcogenide glasses have lower hardness values in general, which is in agreement with [42] which indicates that Ge-Se-based glasses have lower $H_v$ values. Fig 7(c) shows elements like S, Se, Tl, and Ge interacting with most of the elements, together with the violin plot, establishing their major contribution to the $\rho$. The figure also shows the major interaction between S and Se, this is also in accordance with [39]for $\rho$ in chalcogenide glasses. Figure 7(d) indicates the elements above with higher interaction values with other elements. Cu-Fe, Se-Zr, and Se-Fe are the three combinations of elements with the highest interaction value. The Cu-Fe value agrees with the experimental fact of linear dependency of both elements on the *E* value. Figure 7(e) indicates Zr, Se, Ge and Fe, with Zr-Se, Ge-Se, Se-Fe and Zr-Fe are the four combinations of elements with the highest interaction value for *G*. Fig 7(f) indicates Zr, Se, Cu, and Fe interact with most of the elements and has the highest interaction value for $n_d$.

*Glass selection chart*
Due to functional requirements, multiple properties are required in a single glass. However, due to the contrasting effect of elements on desired properties, e.g., the higher concentration of Ge increases the *TEC* but reduces the $T_L$, glass designers need to select chemical constituents such that targeted properties are achieved. Using the glass selection charts (also known as Ashby plots for glasses), functional glass development can be achieved [43], [44][29]. We only consider two-dimensional GSCs since they are easy to visualize, however, multidimensional GSCs can be prepared to screen ChGs for tailored properties.

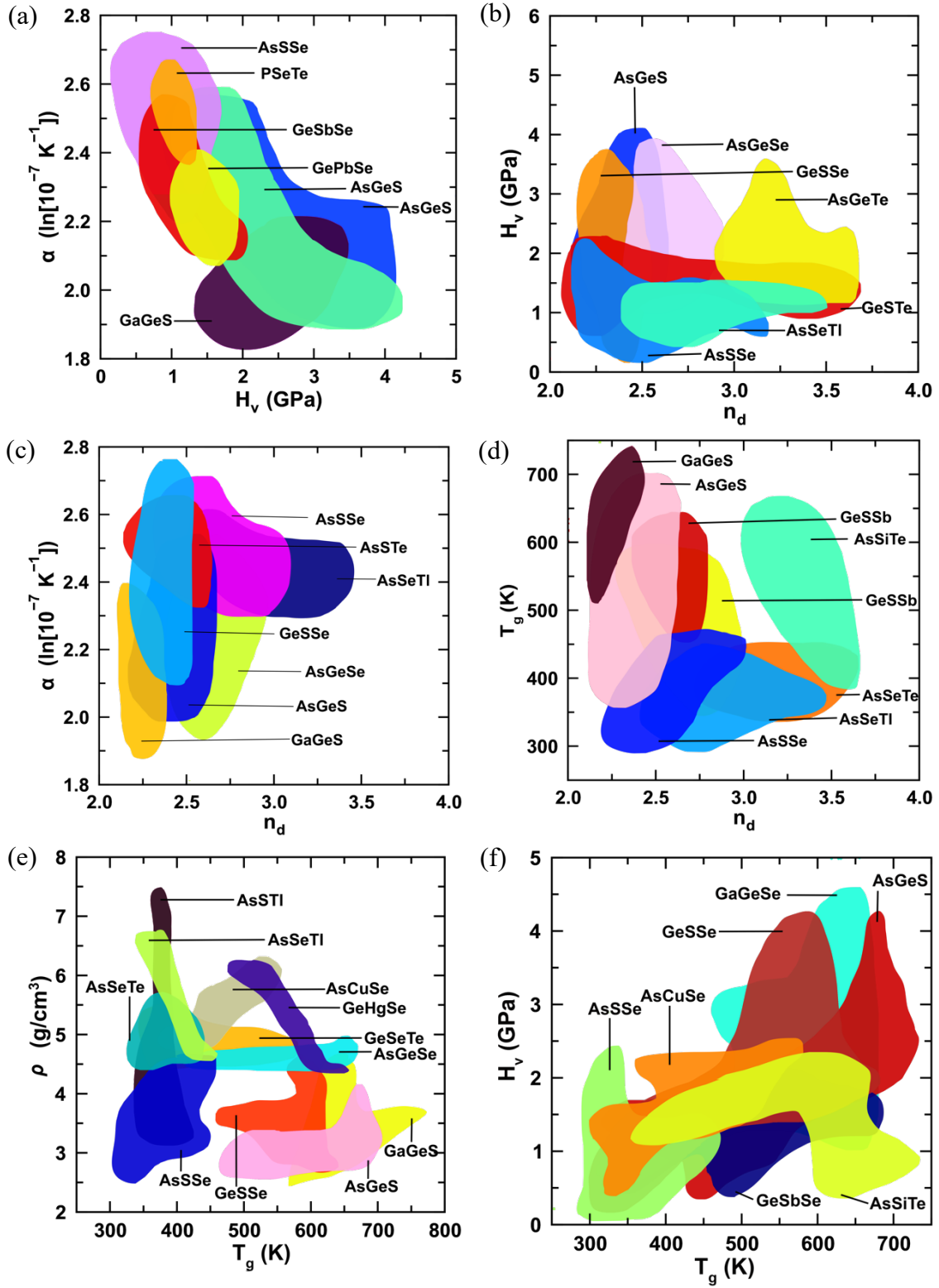

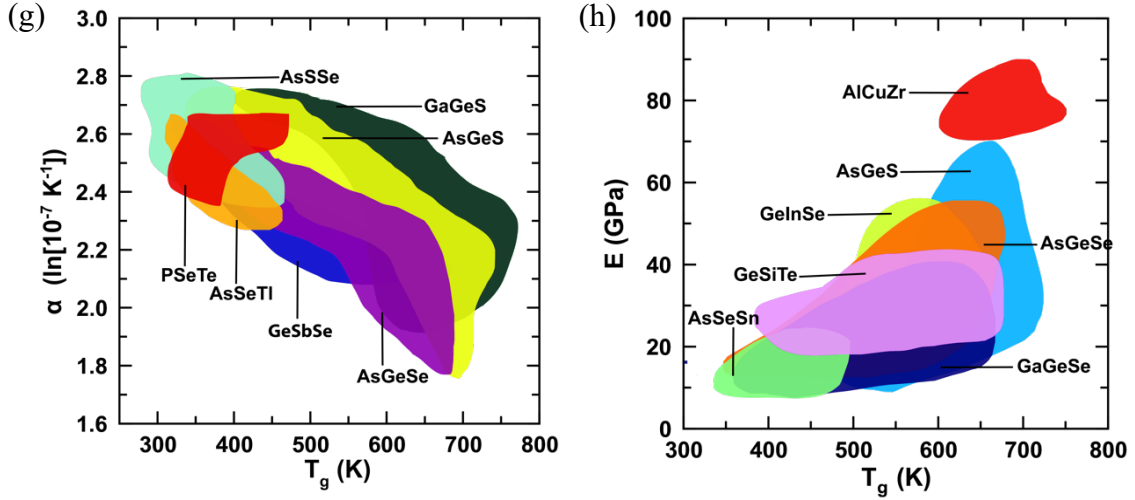

**Figure 8. Glass Selection chart plotted using ML models.** (a) Vicker's hardness ($H_v$) with thermal expansion coefficient (*TEC*), Figs. (b-d) refractive index ($n_d$) with Vickers hardness ($H_v$), thermal expansion coefficient (*TEC*), glass transition temperature ($T_g$), respectively, and Figs. (e-h) represents glass transition temperature ($T_g$) with density ($\rho$),Vickers hardness ($H_v$), thermal expansion coefficient (*TEC*) and young's modulus (*E*), respectively.

For designing optical fibres made of ChGs, it is important to select components to give the desired *TEC* and $H_v$ to facilitate the easy formation of core and clad and deliver desired strength [45], therefore, the GSC presented in Fig. 8(a) can be used. ChGs are known to exhibit phase change phenomena which make them a suitable candidate for temperature sensing application in extreme environments. Figure 8 (b) shows the glass families having high $n_d$ and $H_v$ up to 4.5 GPa. These glasses are used for temperature sensing in nuclear reactors as their $n_d$ changes with a change in the phase from amorphous to crystalline when their temperature rises to crystallization temperature due to radiations in the reactors [46]. Similarly, ChGs for high-stability photonic devices can be selected using Figure 8(c) [47].

To manufacture glasses for high-speed communication fibres and non–linear optical devices, compositions with a wide range of glass transition temperatures dictating the ease of manufacturing of glass and $n_d$ [48] can be obtained from the GSC shown in Fig. 8(d). Glass compositions for nuclear waste immobilization with enhanced $\rho$ and required $T_g$ can be selected using fig.8 (e). ChGs for solid-state electrolytes in Li and Na ion batteries require good mechanical properties and $T_g$ depending upon the range of operational temperatures, hence, Figs. 8(f) can guide the composition selection[49]. The $T_g$ of the material governs their formability, and the *E* of the materials indicates its strength. The glasses belonging to As–Ge–S and As–Ge–Se families feature in both Figs. 8(g) and 8(h). Therefore, glasses for optical devices with required *TEC* and *E* can be obtained using the GSCs presented in Figs. 8(g) and 8(d).

**Conclusions**
Altogether, we showed that ML models can be successfully used to predict twelve important properties of chalcogenide glasses. Further, we showed that the interpretable machine learning method of SHAP analysis can be used to explain the compositional control of these properties of ChGs. Specifically, SHAP bar plots provide the mean absolute effect of each element, while the violin plots explain the effect of the feature with respect to its actual value. Further, SHAP interaction value plots provide insights into how different constituents of the glass interact together and control the properties of the given ChGs. Therefore, the SHAP plots will help the researchers to accelerate the design ChGs with targeted properties by providing insights into the compositional control of the properties. Further, this work also reports the glass selection charts, which enable researchers to select glasses with desired properties. Overall, the insights gained from SHAP plots and GSCs can enable the rational design of chalcogenide glasses in an accelerated fashion.

**ACKNOWLEDGMENTS:**
N. M. Anoop Krishnan acknowledges the financial support for this research provided by the Department of Science and Technology, India, under the INSPIRE faculty scheme (DST/INSPIRE/04/2016/002774) BRNS YSRA (53/20/01/2021-BRNS) and DST SERB Early Career Award (ECR/2018/002228) award by the Government of India. Sayam Singla and Abhishek Vyas acknowledges the funding received from undergraduate research award by IRD unit of IIT Delhi.

**REFERENCES:**

[1] D. Lezal, "Chalcogenide glasses - Survey and progress," *J. Optoelectron. Adv. Mater.*, vol. 5, no. 1, pp. 23–34, Mar. 2003.
[2] A. K. Varshneya, *Fundamentals of Inorganic Glasses*. Elsevier, 2013.
[3] V. F. Kokorina, *Glasses for infrared optics*, vol. 13. CRC press, 1996.
[4] R. Wang and B. Luther-Davies, "Structural and Physical Properties of $Ge_xAs_ySe_{1-x-y}$ Glasses," *Amorphous Chalcogenides: Advances and Applications*, p. 97, 2014.
[5] P. Toupin, L. Brilland, G. Renversez, and J. Troles, "All-solid all-chalcogenide microstructured optical fiber," *Optics Express*, vol. 21, no. 12, pp. 14643–14648, 2013, doi: 10.1364/OE.21.014643.
[6] A. Hayashi and M. Tatsumisago, "Invited paper: Recent development of bulk-type solid-state rechargeable lithium batteries with sulfide glass-ceramic electrolytes," *Electron. Mater. Lett.*, vol. 8, no. 2, pp. 199–207, Apr. 2012, doi: 10.1007/s13391-012-2038-6.
[7] L. Tichý and H. Tichá, "Remark on the glass-forming ability in $GeSe_{1-x}$ and $As_xSe_{1-x}$ systems," *Journal of Non-Crystalline Solids*, vol. 261, no. 1, pp. 277–281, Jan. 2000, doi: 10.1016/S0022-3093(99)00608-0.
[8] R. Bhattoo, S. Bishnoi, M. Zaki, and N. M. A. Krishnan, "Understanding the Compositional Control on Electrical, Mechanical, Optical, and Physical Properties of Inorganic Glasses with Interpretable Machine Learning," *Acta Materialia*, p. 118439, Oct. 2022, doi: 10.1016/j.actamat.2022.118439.
[9] M. Zaki *et al.*, "Interpreting the optical properties of oxide glasses with machine learning and Shapely additive explanations," *J Am Ceram Soc.*, p. jace.18345, Jan. 2022, doi: 10.1111/jace.18345.
[10] S. Bishnoi *et al.*, "Predicting Young's modulus of oxide glasses with sparse datasets using machine learning," *Journal of Non-Crystalline Solids*, vol. 524, p. 119643, Nov. 2019, doi: 10.1016/j.jnoncrysol.2019.119643.
[11] S. Bishnoi, R. Ravinder, H. Singh Grover, H. Kodamana, and N. M. Anoop Krishnan, "Scalable Gaussian processes for predicting the optical, physical, thermal, and mechanical properties of inorganic glasses with large datasets," *Materials Advances*, 2021, doi: 10.1039/D0MA00764A.
[12] R. Ravinder *et al.*, "Deep learning aided rational design of oxide glasses," *Materials Horizons*, 2020, doi: 10.1039/D0MH00162G.

[13] E. Alcobaça *et al.*, "Explainable Machine Learning Algorithms For Predicting Glass Transition Temperatures," *Acta Materialia*, vol. 188, pp. 92–100, Apr. 2020, doi: 10.1016/j.actamat.2020.01.047.
[14] D. R. Cassar, "ViscNet: Neural network for predicting the fragility index and the temperature-dependency of viscosity," *Acta Materialia*, vol. 206, p. 116602, Mar. 2021, doi: 10.1016/j.actamat.2020.116602.
[15] D. R. Cassar, G. G. Santos, and E. D. Zanotto, "Designing optical glasses by machine learning coupled with a genetic algorithm," *Ceramics International*, vol. 47, no. 8, pp. 10555–10564, Apr. 2021, doi: 10.1016/j.ceramint.2020.12.167.
[16] D. R. Cassar, A. C. P. L. F. de Carvalho, and E. D. Zanotto, "Predicting glass transition temperatures using neural networks," *Acta Materialia*, vol. 159, pp. 249–256, Oct. 2018, doi: 10.1016/j.actamat.2018.08.022.
[17] M. Zaki, Jayadeva, and N. M. A. Krishnan, "Extracting processing and testing parameters from materials science literature for improved property prediction of glasses," *Chemical Engineering and Processing - Process Intensification*, p. 108607, Aug. 2021, doi: 10.1016/j.cep.2021.108607.
[18] Ravinder *et al.*, "Artificial intelligence and machine learning in glass science and technology: 21 challenges for the 21st century," *International Journal of Applied Glass Science*, vol. n/a, no. n/a, 2021, doi: https://doi.org/10.1111/ijag.15881.
[19] C. Dreyfus and G. Dreyfus, "A machine learning approach to the estimation of the liquidus temperature of glass-forming oxide blends," *Journal of Non-Crystalline Solids*, vol. 318, no. 1, pp. 63–78, Apr. 2003, doi: 10.1016/S0022-3093(02)01859-8.
[20] T. Gupta, M. Zaki, N. M. A. Krishnan, and Mausam, "MatSciBERT: A materials domain language model for text mining and information extraction," *npj Comput Mater*, vol. 8, no. 1, p. 102, Dec. 2022, doi: 10.1038/s41524-022-00784-w.
[21] M. Zaki *et al.*, "Natural language processing-guided meta-analysis and structure factor database extraction from glass literature," *Journal of Non-Crystalline Solids: X*, vol. 15, p. 100103, Sep. 2022, doi: 10.1016/j.nocx.2022.100103.
[22] J. C. Mauro, A. Tandia, K. D. Vargheese, Y. Z. Mauro, and M. M. Smedskjaer, "Accelerating the Design of Functional Glasses through Modeling," *Chemistry of Materials*, vol. 28, no. 12, pp. 4267–4277, Jun. 2016, doi: 10.1021/acs.chemmater.6b01054.
[23] S. M. Mastelini, D. R. Cassar, E. Alcobaça, T. Botari, A. C. P. L. F. de Carvalho, and E. D. Zanotto, "Machine learning unveils composition-property relationships in chalcogenide glasses," *Acta Materialia*, vol. 240, p. 118302, Nov. 2022, doi: 10.1016/j.actamat.2022.118302.
[24] S. M. Mastelini, D. R. Cassar, E. Alcobaça, T. Botari, A. C. P. L. F. de Carvalho, and E. D. Zanotto, "Machine learning unveils composition-property relationships in chalcogenide glasses," *arXiv:2106.07749 [cond-mat]*, Jun. 2021, Accessed: Jun. 22, 2021. [Online]. Available: http://arxiv.org/abs/2106.07749
[25] S. Ramraj, N. Uzir, R. Sunil, and S. Banerjee, "Experimenting XGBoost algorithm for prediction and classification of different datasets," *International Journal of Control Theory and Applications*, vol. 9, no. 40, 2016.
[26] T. Akiba, S. Sano, T. Yanase, T. Ohta, and M. Koyama, "Optuna: A Next-generation Hyperparameter Optimization Framework," *arXiv:1907.10902 [cs, stat]*, Jul. 2019, Accessed: Oct. 28, 2021. [Online]. Available: http://arxiv.org/abs/1907.10902
[27] J. Bergstra, R. Bardenet, Y. Bengio, and B. Kégl, "Algorithms for hyper-parameter optimization," *Advances in neural information processing systems*, vol. 24, 2011.
[28] N. Hansen and A. Ostermeier, "Completely derandomized self-adaptation in evolution strategies," *Evolutionary computation*, vol. 9, no. 2, pp. 159–195, 2001.
[29] B. Shahriari, K. Swersky, Z. Wang, R. P. Adams, and N. De Freitas, "Taking the human out of the loop: A review of Bayesian optimization," *Proceedings of the IEEE*, vol. 104, no. 1, pp. 148–175, 2015.

[30] K. Jamieson and A. Talwalkar, "Non-stochastic best arm identification and hyperparameter optimization," presented at the Artificial intelligence and statistics, 2016, pp. 240–248.

[31] A. Gorbunov, S. Liav, and A. Tverjanovich, "Thermal expansion coefficient and relaxation parameters of glasses in the system GaX2-GeX2-Sb2X3 (X= S, Se)," *Glass Physics and Chemistry*, vol. 38, no. 3, pp. 269–273, 2012.

[32] J. D. Musgraves, S. Danto, and K. Richardson, "Thermal properties of chalcogenide glasses," in *Chalcogenide Glasses*, Elsevier, 2014, pp. 82–112.

[33] V. S. Shiryaev, A. I. Filatov, E. V. Karaksina, and A. V. Nezhdanov, "Structural peculiarities of Ge-rich Ga-Ge-Sb-Se chalcogenide glasses," *Journal of Solid State Chemistry*, vol. 303, p. 122454, Nov. 2021, doi: 10.1016/j.jssc.2021.122454.

[34] P. K. Thiruvikraman, "Rings, chains and planes: Variation of $T_g$ with composition in chalcogenide glasses," *Bull Mater Sci*, vol. 29, no. 4, pp. 371–374, Aug. 2006, doi: 10.1007/BF02704137.

[35] U. Senapati and A. K. Varshneya, "Configurational arrangements in chalcogenide glasses: A new perspective on Phillips' constraint theory," *Journal of non-crystalline solids*, vol. 185, no. 3, pp. 289–296, 1995.

[36] A. E. Voronova, V. A. Ananichev, and L. N. Blinov, "Thermal expansion of melts and glasses in the As–Se system," *Glass physics and chemistry*, vol. 27, no. 3, pp. 267–273, 2001.

[37] K. Samudrala and S. B. Devarasetty, "Effect of the Copper on Thermo - Mechanical and Optical Properties of S-Se-Cu Chalcogenide Glasses," *IOP Conference Series. Materials Science and Engineering*, vol. 330, no. 1, Mar. 2018, doi: 10.1088/1757-899X/330/1/012042.

[38] J. Cui *et al.*, "The influence of different antimony content in Ga-As-Sb-S chalcogenide glass system: Modification of physical & spectroscopic properties and fiber forming ability," *Ceramics International*, 2022.

[39] Y. Yang *et al.*, "Composition dependence of physical and optical properties in Ge-As-S chalcogenide glasses," *Journal of Non-Crystalline Solids*, vol. 440, pp. 38–42, May 2016, doi: 10.1016/j.jnoncrysol.2016.03.003.

[40] Z. Cao *et al.*, "Chalcogenide glass ceramics: A high-performing innovative infrared acousto-optic material," *Journal of the European Ceramic Society*, vol. 41, no. 14, pp. 7215–7221, 2021.

[41] L. G. Aio, A. M. Efimov, and V. F. Kokorina, "Refractive index of chalcogenide glasses over a wide range of compositions," *Journal of Non-Crystalline Solids*, vol. 27, no. 3, pp. 299–307, Mar. 1978, doi: 10.1016/0022-3093(78)90015-7.

[42] J.-P. Guin, T. Rouxel, J.-C. Sanglebœuf, I. Melscoët, and J. Lucas, "Hardness, Toughness, and Scratchability of Germanium–Selenium Chalcogenide Glasses," *Journal of the American Ceramic Society*, vol. 85, no. 6, pp. 1545–1552, 2002, doi: 10.1111/j.1151-2916.2002.tb00310.x.

[43] M. F. Ashby, Y. J. M. Bréchet, D. Cebon, and L. Salvo, "Selection strategies for materials and processes," 2004. Accessed: Oct. 26, 2022. [Online]. Available: https://www.sciencedirect.com/science/article/pii/S0261306903001596

[44] M. F. Ashby and K. Johnson, *Materials and Design: The Art and Science of Material Selection in Product Design*. Butterworth-Heinemann, 2013.

[45] J. S. Sanghera and I. D. Aggarwal, "Active and passive chalcogenide glass optical fibers for IR applications: a review," *Journal of Non-Crystalline Solids*, vol. 256–257, pp. 6–16, Oct. 1999, doi: 10.1016/S0022-3093(99)00484-6.

[46] A.-A. A. Simon, K. Kadrager, B. Badamchi, H. Subbaraman, and M. Mitkova, "Temperature sensing in nuclear facilities: Application of the phase change effect of chalcogenide glasses," *Proceedings of the Nuclear Plant Instrumentation, Control, and Human-Machine Interface Technologies, Orlando, FL, USA*, pp. 9–14, 2019.

[47] L. Zhu *et al.*, "Optical and thermal stability of Ge-as-Se chalcogenide glasses for femtosecond laser writing," *Optical Materials*, vol. 85, pp. 220–225, Nov. 2018, doi: 10.1016/j.optmat.2018.08.041.
[48] P. Kumar, J. Kaur, S. K. Tripathi, and I. Sharma, "Effect of antimony (Sb) addition on the linear and non-linear optical properties of amorphous Ge–Te–Sb thin films," *Indian J Phys*, vol. 91, no. 12, pp. 1503–1511, Dec. 2017, doi: 10.1007/s12648-017-1053-8.
[49] Y. Zhang, Q. Jiao, B. Ma, C. Lin, X. Liu, and S. Dai, "Structure and ionic conductivity of new Ga2S3-Sb2S3-NaI chalcogenide glass system," *Physica B: Condensed Matter*, vol. 570, pp. 53–57, Oct. 2019, doi: 10.1016/j.physb.2019.05.026.